\documentstyle[multicol,prl,aps,tighten]{revtex}

\begin{document}

\draft

\title{Frozen Disorder in a Driven System}
\author{B. Schmittmann$^{1}$ and K. E. Bassler$^{1,2,*}$}
\address{$^{1}$Center for Stochastic Processes in Science and Engineering\\
and Department of Physics, Virginia Tech, Blacksburg, VA 24061-0435\\
$^{2}$Department of Physics and Astronomy, 
Louisiana State University,
Baton Rouge, LA 70803}

\date{February 21, 1996; revised September 9, 1996}
\maketitle

\begin{abstract}
We investigate the effects of quenched disorder on the universal
properties of a randomly driven Ising lattice gas. The Hamiltonian fixed
point of the pure system becomes unstable in the presence of a quenched
local bias, giving rise to a new fixed point which controls a novel
universality class. We determine the associated scaling forms of correlation
and response functions, quoting critical exponents to two-loop order in 
an expansion around the upper critical dimension d$_c=5$. 
\end{abstract}

\begin{multicols}{2}
\narrowtext
The statistical mechanics of systems which settle into non-equilibrium
steady states (NESS) has attracted considerable interest recently 
\cite{review}.
Much attention focused on Ising-like systems subject to
a driving force. These models exhibit many intriguing features, such as
anomalous correlations at all temperatures or order-disorder transitions
belonging into various non-equilibrium universality classes \cite{review}.
Yet their basic building blocks are designed to be as simple as possible, 
so
that they are easily simulated {\em and} accessible by analytic methods.
Thus, key characteristics of generic non-equilibrium behavior can be traced
to specific model ingredients, a procedure which would be excessively
difficult if one attempted to capture the full complexity of a real system.

Microscopic models of driven systems typically involve nearest-neighbor
particle-hole exchanges on a fully periodic regular lattice, according 
to a
sequential dynamics which is local and translation-invariant, in
space
and time. The hopping rates are controlled solely by (i) differences in
internal energy, determined by a local Hamiltonian, (ii) the coupling 
to a
heat bath which enters through its temperature T, and (iii) a driving
force. The latter acts as a dynamic perturbation which drives the system
away from thermal equilibrium, into a non-Hamiltonian steady state which
depends on the details of the dynamics. The prototype is a {\em uniformly}
driven system \cite{kls}, motivated by the
physics of fast ionic conductors \cite{fic's}: Here, the usual Ising
energetics is modified by a uniform bias which favors (suppresses) jumps
along (against) a specific lattice direction, 
thus inducing a global current
through the system. Recast as a Langevin equation in continuous space 
and
time, the theory can be analyzed using mean-field or
renormalization group methods. Since, in principle, 
this procedure relies on
coarse-graining the microscopic dynamics, the resulting equation is
identical for all choices of {\em local} microscopic rules 
which respect the
same set of global symmetries and conservation laws. In contrast, different
universal behavior, controlled by a modified Langevin equation, is
expected if the dynamics is {\em nonlocal}, in space or time.

In this letter, we consider the effect of nonlocality in time. In any
experimental realization of driven lattice gases, the presence of disorder,
in the form of impurities or lattice defects, is to be expected. If such
defects are effectively frozen on the time scales that govern the
the order parameter dynamics, they introduce long (technically infinite)
correlation times, or memory effects, into the dynamics, which may lead 
to
striking consequences for static and dynamic critical behavior. In
equilibrium systems, the study of quenched disorder has a long history 
\cite{stinch}, and some general features, 
such as the Harris criterion \cite{harris} 
for the stability of pure fixed points, can be identified. For
driven systems, given the absence of a free energy, it is not known whether
any such simple criteria exist. We therefore present a case study of a
specific driven system, as a step towards a more general exploration
of the effects of quenched randomness on NESS.

The choice of our model is motivated by two earlier studies on frozen
disorder in the {\em uniformly} driven lattice gas, one simulational \cite
{lauritsen} and the other analytic \cite{bec+jan2}, considering,
respectively, a small concentration of immobile impurities versus quenched
randomness in the nearest-neighbor hopping rates \cite{bec+jan1}. In both
cases, thermodynamic observables are averages over different
disorder realizations. 
However, the presence of a global current, induced by the uniform
drive, leads to a fundamental difficulty: by necessity,
the {\em simulations} are performed on a finite periodic lattice,
and before the system reaches steady state, individual particles
have been driven through the lattice several times, encountering the {\em 
same}
disorder configuration on each visit. 
Thus,
the usual time averages include uncontrolled correlations in the
randomness, in stark contrast to the physics of real systems,
where particles travel through non-repeating impurity environments.
Also, the renormalization group predictions are
predicated upon uncorrelated disorder, so that comparisons 
are not possible.

Clearly, this predicament presents a serious drawback for a baseline study
in novel territory. We are thus motivated to undertake 
the
first case study which is free from this difficulty, by investigating 
the
effects of frozen disorder on a NESS {\em without} global current. In the
following, we first describe our microscopic model and derive
the associated Langevin equation. Next, we compute its universal scaling
properties to two-loop order in an expansion around the upper critical
dimension $d_c=5$. To illustrate our predictions, we consider
the structure factor, since the latter is easily measured 
in
simulations. We conclude with some comments and open questions.

In the absence of quenched disorder, our model \cite{sch+zia1} consists 
of
particles and holes,
distributed over a fully periodic regular lattice, with the usual
nearest-neighbor attractive Ising interaction. 
The drive $\vec{E}$ $=E(x,t)\hat{e}$ acts 
along a specific lattice direction $\hat{e}$, labelled
`parallel', and is characterized by an {\em annealed} random amplitude, 
$E(x,t)$. Particle-hole exchanges in the transverse directions occur
according to the usual Ising energetics, 
while particle hops along (against) 
$\vec{E}$ are enhanced (suppressed) by a factor of 
$\exp(+E/T)$ ( $\exp(-E/T)$). 
For each parallel exchange, occurring at time $t$ and lattice site $x$, 
a
new value of $E(x,t)$ is chosen from a symmetric distribution with zero
mean, so that no global current through the system is induced. Otherwise,
the detailed form is irrelevant for universal critical behavior.

Like its uniformly driven counterpart, this randomly driven system exhibits
a continuous transition, from a disordered into an strip-like ordered 
phase.
It is in the same universality class as an Ising lattice gas with Kawasaki
dynamics, controlled by two temperatures \cite{kett}. Its scaling properties
have been calculated to two-loop order in an expansion around its an upper
critical dimension $d_c=3$ \cite{sch+zia1,sch}, and agree well with
results from simulations \cite{praest}. Intriguingly, the {\em fixed point}
of this non-equilibrium system is
Hamiltonian \cite{sch}, formally identical to the one controlling the 
{\em equilibrium} scaling behavior of uniaxial ferromagnets with long-range
dipolar interactions \cite{dipolar}.

Next, we introduce frozen disorder into the {\em dynamics} of the system, 
by adding a quenched random contribution, $\Delta (x)$, 
to the amplitude of the drive. 
Thus, we focus on the non-equilibrium aspects of the problem, in
particular since the {\em statics} of uniaxial ferromagnets with quenched
disorder has already been discussed elsewhere \cite{aha}. Writing 
$\vec{E}=[E(x,t)+\Delta (x)]\hat{e}$, 
we assume that the quenched component $\Delta
(x)$ is distributed symmetrically around zero mean, so that no global
current is induced, with second moment 
$\langle \Delta (x)\Delta (x^{\prime})\rangle 
=2\sigma _{\parallel }\,\delta (x-x^{\prime })$. Note that there
is no $\delta $-function in time here, consistent with the quenched
character of this term. Again, the detailed form of the distribution is
irrelevant for leading singular behavior. This type of randomness is easily
implemented in a Monte Carlo simulation, where it corresponds to a
time-independent local bias resident on each longitudinal bond. In the
absence of $\Delta $, the system reduces to the randomly driven (or
two-temperature) lattice gas which we will refer to as the `pure' system. 

The only slow variable of our theory is the local magnetization, 
$\phi (\vec{x},t)$. Since it is conserved, 
the effective equation of motion for its
long-wavelength, long-time behavior takes the form of a continuity equation, 
$\partial _t\phi =-\vec{\nabla}\cdot \vec{j}$. The current $\vec{j}$\ 
here
is composed of two terms, a deterministic and a noisy part. In the absence
of $\vec{E}$, it is given by Model B of the Halperin-Hohenberg scheme 
\cite
{hal+hoh}, i.e., 
$\vec{j}=-\vec{\nabla}(\delta {\cal H}/\delta \phi )+\vec{\zeta}$, 
where ${\cal H[}\phi ]\;$is the usual Landau-Ginzburg-Wilson
Hamiltonian. $\vec{\zeta}$ represents the {\em thermal} noise, modeling 
the
relevant effects of the fast microscopic degrees of freedom after
coarse-graining. It is Gaussian distributed with zero mean 
and second moment 
$\langle \zeta _i(\vec{x},t)\zeta _j(\vec{x},t^{\prime })\rangle 
\!=2\lambda
N_{ij}\delta (x-x^{\prime })\delta (t-t^{\prime })$. 
Here, $i,j$ label the
cartesian coordinates of $\vec{\zeta}$, 
and $N_{ij}$ is a noise correlation
matrix. In equilibrium, the 
fluctuation-dissipation theorem (FDT) \cite{kubo}
requires $N_{ij}\propto \delta _{ij}$. A non-vanishing $\vec{E}$ has two
major effects on the equation of motion:\ First, it induces an additional
contribution, $\vec{j}_E$, to the deterministic part of $\vec{j}$. Familiar
from the uniformly driven system, $\vec{j}_E=c(\phi )\vec{E}$ is an Ohmic
term, with the conductivity $c(\phi )\propto 1-\phi ^2$ reflecting the
vanishing of $\vec{j}_E$ in a completely filled or empty system. Second,
since $\vec{E}$ singles out a preferred direction, $\vec{j}$\/\ will become
anisotropic, so that all couplings have to be replaced by appropriate
anisotropic ones \cite{sch+zia1}. In particular, 
there will be two diffusion
coefficients, $\tau _{\parallel }$ \ and $\tau _{\perp }$. 
Given that $\vec{E}$ 
acts like an extra noise along the parallel direction, $\tau _{\parallel}$ 
is enhanced over $\tau _{\perp }$, so that criticality is associated
with $\tau _{\perp }$ $\rightarrow 0$ while $\tau _{\parallel }$ \ remains
positive. Moreover, while still diagonal, the noise correlation matrix 
is 
no
longer proportional to the unit matrix. Instead, $\vec{\zeta}(\vec{x},t)$
splits into a parallel component, $\zeta _{\parallel }$ , and a 
$(d-1)$-dimensional transverse component, $\vec{\zeta}_{\perp }$, 
distributed
according to $\exp \left[ -\int d^dxdt\left( \zeta _{\parallel
}^2/4n_{\parallel }+\zeta _{\perp }^2/4n_{\bot }\right) \right] $, so 
that
the FDT is broken. Finally, we return to the divergence of the Ohmic
current, $\vec{\nabla} \cdot \vec{j}_E \propto \vec{\nabla}
\cdot \left[ (1-\phi ^2)\vec{E} \right]$,
and insert $\vec{E} = [E(x,t)+\Delta (x)]\hat{e}$. Letting $\nabla $ ($
\partial $) denote the gradient in the transverse (longitudinal) directions,
we recognize that the term $\partial E(x,t)$ simply enhances the thermal
noise $\partial \zeta _{\parallel }$ . In contrast, $\partial \Delta (x)$
gives rise to a new term, corresponding to a {\em quenched random} Langevin
noise, which will be responsible for profound changes in the critical
behavior, compared to the pure system. The remaining term, $\partial \phi
^2(E+\Delta )$, generates only couplings that are irrelevant in the
renormalization group sense.

The resulting Langevin equation of motion must, in principle,
be solved for $\phi (x,t)$, so that thermodynamic
observables, such as correlation functions, can be computed by averaging
over the thermal noise $\vec{\zeta}$, the annealed drive $E(x,t)$ and 
the
frozen disorder $\Delta (x)$. In practice, it is easier to
introduce a Martin-Siggia Rose \cite{MSR} response field $\tilde{\phi}$, 
and
recast the Langevin equation as a dynamic functional \cite{dyn fun}, 
${\cal J}[\phi ,\tilde{\phi}]$. In this form, the averages over 
$\vec{\zeta}$, $E(x,t)$ and $\Delta (x)$ 
are easily performed, and
correlation and response functions 
follow as functional averages weighted by 
$\exp (-{\cal J})$. Neglecting irrelevant terms,
\end{multicols}
\widetext
\begin{eqnarray}
{\cal J\,}[\phi ,\tilde{\phi}] &=&\int d^dx\int dt\tilde{\phi}\left\{ 
\dot{\phi}-\lambda \left[ (\tau _{\perp }-\nabla ^2)\nabla ^2\phi 
+\tau_{\parallel }\partial ^2\phi +\frac g{3!}\nabla ^2\phi ^3\right] 
\right\}  
\nonumber \\
& &+\int d^dx\{\int dt\lambda \tilde{\phi}(x,t)(n_{\parallel }\partial^2
+n_{\perp }\nabla ^2)\tilde{\phi}(x,t)
+\int dt\int dt^{\prime }\lambda^2\tilde{\phi}(x,t)(\sigma_{\parallel 
}
\partial^2
-\sigma _{\perp }\nabla^4)
\tilde{\phi}(x,t^{\prime})\}  \label{1}
\end{eqnarray}
\begin{multicols}{2}
\narrowtext
The term in the square brackets is easily recognized as the deterministic
part of $\vec{\nabla}\cdot \vec{j}.$ The two contributions that are
quadratic in $\tilde{\phi}$ arise from the averages over the thermal noise 
$\vec{\zeta}(x,t)$ and the quenched disorder $\Delta (x)$, respectively.
Since $\vec{\zeta}$ is $\delta $-correlated in {\em both }space and
time, the resulting operator is fully local. In contrast, the operator
associated with the quenched disorder is delocalized in {\em time},
reflecting the infinite correlation time of $\Delta (x)$. We emphasize 
that it
consists of two terms: In addition to the coupling $\sigma _{\parallel 
}$ ,
arising from the average over $\Delta (x)$, a second time-delocalized
coupling $\sigma _{\perp }$ is generated under the renormalization group.
It is required for closure and must be included in (1), even though it 
did not
appear in our original Langevin equation. Finally,
we note that all coefficients here are functions of the microscopic model
parameters, but - in the spirit of Landau theory - their explicit dependence
need not be known.

The first step towards the renormalization group analysis of our theory
consists in dimensional analysis. Introducing a typical scale $\mu $ to
measure transverse momenta $k_{\perp }$, we first note that, near
criticality, parallel momenta scale as $k_{\parallel }\propto \mu^2$. 
Next,
requiring that all terms in (1) be dimensionless,
we find $\lambda t\propto \mu^{-4}$,
so that the thermal noise becomes {\em irrelevant }and may be neglected,
compared to the noise term associated with the frozen disorder. 
Furthermore, choosing the scale of $\tilde{\phi}$ such that 
$\sigma_{\perp} \propto \mu^0$, we find
$\phi \propto \mu^{(d-3)/2}$ and $\tilde{\phi} \propto \mu^{(d+5)/2}$,
implying that $g \propto \mu^{5-d}$.
The upper
critical dimension follows as $d_c=5$, which is shifted upwards 
from pure system where $d_c=4$. Thus, our first major result emerges, 
namely, that the
disordered system belongs to a different
universality class than the pure system. 
Similar dimensional shifts have been observed in the
uniformly driven system when frozen disorder is included 
\cite{bec+jan2,bec+jan1}. However, we note that the 
$\varepsilon$-expansions of the
disordered and the pure systems do not coincide here, in contrast to
random-field ferromagnets in equilibrium \cite{rff}.

Following standard methods \cite{standard FT stuff}, we now consider the
critical theory with insertions of 
$\lambda \tau _{\perp }\tilde{\phi}\nabla^2\phi $, to two-loop order in 
$\varepsilon \equiv d-d_c$. The bare
correlation and response propagators take the anisotropic forms dictated 
by
(1), which differ from their counterparts in Model B or the pure system. 
In
particular, the correlator $S(k,\omega )$ becomes proportional to $\delta
(\omega )$, by virtue of the time-delocalized noise. 
Noting that (1) is invariant under the anisotropic scale
transformation $x_{\parallel }\rightarrow \alpha x_{\parallel }$, 
$x_{\perp}\rightarrow x_{\perp }$, 
$\tilde{\phi}\rightarrow \beta \tilde{\phi}$, 
$\phi \rightarrow (\alpha \beta )^{-1}\phi $, 
the effective dimensionless expansion parameters of the theory 
can be identified as $v\equiv
\sigma _{\perp }\tau _{\parallel }^{-1/2}g$, and 
$w\equiv \sigma _{\parallel}/(\sigma _{\perp }\tau _{\parallel })$.

Next, we proceed to compute the one-particle irreducible vertex functions
with $n$ ($\tilde{n}$) external $\phi $- ($\tilde{\phi}$-) legs and $m$
insertions, denoted by $\Gamma _{\tilde{n}n}^{(m)}$, in a double expansion
in $v$ and $w$. As in the pure system, $\Gamma _{11}^{(0)}$, $\Gamma
_{13}^{(0)}$, as well as $\Gamma _{11}^{(1)}$ are ultraviolet divergent 
and
can be rendered finite by renormalizing $\phi $, $g$, and $\tau _{\perp 
}$,
respectively. In addition, however, $\Gamma _{20}^{(0)}$ {\em also} requires
renormalization here, by redefining $\sigma _{\perp }$. Beyond these four,
no other independent renormalizations are needed. For $\varepsilon >0$, 
we
find exactly one stable, nontrivial fixed point $(v^{*},w^{*})$, which
controls the critical behavior of the theory below five dimensions.
Surprisingly, the coupling $w$ is found to be {\em irrelevant} under the 
RG,
approaching its fixed point value $w^{*}=0$ quite slowly, with a
correction-to-scaling exponent $\omega _w=\frac 4{81}\varepsilon ^2\left\{
\ln \frac 43-\frac{13}{54}\right\} +O(\varepsilon ^3)$. $v$, on the other
hand, flows much faster to a finite value 
$v^{*}=\frac 89\varepsilon +(\frac 89)^3C\varepsilon ^2+O(\varepsilon 
^3)$, 
controlled by a correction-to-scaling exponent 
$\omega _v=\varepsilon -(\frac
89)^2C\varepsilon ^2+O(\varepsilon ^3)$. Here, 
$C=\frac{5129}{3^32^7}+\frac 72\ln 2-\frac{29}{32}\ln 3$ 
is a numerical constant.

Incorporating dimensional analysis and the scale invariance of the theory,
the solution of the renormalization group equations at the fixed point
yields the full scaling behavior of, e.g., 
the steady-state structure factor,
\begin{equation}
S(k,t)=\mu ^{-4+\eta }S(k_{\parallel }/\mu ^{1+\Delta },
k_{\perp }/\mu ,t\mu^z;\tau _{\perp }/\mu ^{1/\nu })  
\label{2}
\end{equation}
which serves to define the four indices $\nu $, $\eta $, $z$, and $\Delta 
$ 
\cite{review}. In particular, we find that $\nu =\frac 12+\frac \varepsilon
{12}+\frac{\varepsilon ^2}{108}\left[ \frac{1474}{243}+\frac{31}9\ln \frac
43+2\ln 4\right] +O(\varepsilon ^3)$ is determined by the 
renormalization of 
$\tau _{\perp }$, resulting in an exponent that is distinct from its
counterpart in the pure system \cite{sch}. The remaining two independent
renormalizations, of the field $\phi $ and the coupling $\sigma _{\perp 
}$,
give rise to two independent exponents, $\bar{\eta}\equiv \frac{37}{2187}
\varepsilon ^2+O(\varepsilon ^3)$ and 
$\bar{\sigma}\equiv \frac 4{243} \varepsilon^2 \left[ \frac{11}{36}
+3\ln \frac 43\right] +O(\varepsilon^3)$. 
Scaling laws express the {\em three} exponents $\eta$, $z$, and $\Delta 
$ in terms of 
these two quantities. Specifically, the dynamic
exponent $z$ follows as $z=4-\bar{\eta}$ while the strong anisotropy
exponent $\Delta $ is given by $\Delta =1-\bar{\eta}/2$. Thus, we recover 
a
scaling law, $z=4-2(1-\Delta )$, which is also obeyed by the pure system. 
In contrast, however, the exponent $\eta $ is {\em not} simply related 
to 
$\bar{\eta}$: 
Unlike the pure system, where $\eta =\bar{\eta}$, we find 
$\eta =2- \bar{\sigma}
=\frac{\varepsilon^2}{81}\left[ \frac 73-4\ln \frac
43\right] +O(\varepsilon ^3)$ by virtue of the additional independent
renormalization of the coupling $\sigma _{\perp }$.

Other scaling laws determine the remaining anisotropic exponents \cite
{review}. Some of these involve only $\Delta $, and hence $\bar{\eta}$, 
such as the two different correlation length indices 
$\nu _{\perp }=\nu $ and 
$\nu _{\parallel }=\nu (1+\Delta )$, or the two different dynamic exponents,
$z_{\perp }=z$ and $z$ $_{\parallel }=z/(1+\Delta )$. Since $\Delta >0$, 
we
find $z_{\parallel}< z_{\perp}$ so that transverse critical
fluctuations decay more slowly than parallel ones. On the other hand, 
the
four $\eta $-like exponents, characterizing the anisotropic power-law
behavior of the structure factor in real (RS) and momentum (MS) space, 
are
determined by{\em \ both} $\eta $ {\em and} $\bar{\eta}$: $\eta _{\perp
}^{MS}=\eta $, $\eta _{\parallel }^{MS}=(\eta +2\Delta )/(1+\Delta )$, 
$\eta_{\perp }^{RS}=\eta +\Delta $, 
and $\eta _{\parallel }^{RS}=[\eta-(d-3)\Delta ]/(1+\Delta )$. 
Finally, we quote the order parameter exponent 
$\beta =\frac 12\nu (d-4+\eta +\Delta )$ 
which follows from the scaling form
of the equation of state \cite{standard FT stuff}.

In summary, we have analyzed the critical behavior of a randomly driven
stochastic lattice gas, under the additional influence of a frozen local
bias, acting along the parallel direction. This bias produces a
time-delocalized Langevin current, with components in both, parallel and
transverse, directions, that dominates the usual thermal noise. 
As a result, the
critical dimension is shifted upwards to $d_c=5$, and the critical behavior
falls into a new universality class, distinct from the pure system.

We conclude with two remarks. First, we return briefly to the microscopic
mechanisms giving rise to the coupling $\sigma _{\perp }$, i.e., the
transverse part of the frozen Langevin current. As discussed above,
it is generated by a quenched bias acting along parallel bonds.
Intriguingly, however, the {\em same} term also arises in a very different
microscopic context, namely, if the quenched disorder resides in the
nearest-neighbor hopping rates \cite{bec+jan1}, modelled, e.g., by a
Gaussian distribution of local barriers between sites. Characterizing 
this
landscape by a coarse-grained potential \cite{review} $U(x)$, the Langevin 
{\em current} takes the form of a potential gradient, $\nabla U(x)$.
Averaging over $U$ then leads to $\int dt\int dt^{\prime }\lambda ^2
\tilde{\phi}(x,t)(\sigma _{\perp }\nabla ^4
+\kappa _{\parallel }\partial ^4)\tilde{\phi}(x,t^{\prime })$ 
in (1) instead, with the latter term being irrelevant
for critical behavior near $d_c=5$. Thus, two rather different microscopic
models, distinguished by the implementation of the frozen disorder, are
described by the same mesoscopic Langevin equation, so that they both 
fall
into the same universality class. On the other hand, a quenched bias acting
along {\em transverse} bonds gives rise to 
different universal behavior \cite
{sch+lab}. Work is in progress to clarify the 
intriguing connections between
different microscopics with frozen disorder and their surprising degree 
of
universality. Second, we wish to comment on the
observability of our predictions in a computer experiment. To measure 
the
full time-dependent structure factor, 
the leading singular part, captured by (2), 
must be separated from corrections to scaling, 
generated by irrelevant couplings such as, e.g., the
thermal noise terms. Since the latter decay exponentially in time, only 
the
former survives for large t. Its prime signature is its strong 
$k_{\perp}^{-4+\eta }$ divergence as $k_{\perp }\rightarrow 0$. 
Clearly, it would
be very interesting to test our predictions with
simulations.

We wish to thank V. Becker, H.K. Janssen, K. Oerding and R.K.P. Zia for
many helpful discussions. We are particularly grateful to V. Becker
and H.K. Janssen for communicating the results of Ref.~\cite{bec+jan2} 
prior to publication. 
This work is supported in part by the National Science
Foundation through DMR-9419393 and DMR-9408634.

\end{multicols}

\end{document}